\documentclass[preprint,showpacs,preprintnumbers,amsmath,amssymb]{revtex4}

\usepackage{graphicx}
\usepackage{dcolumn}
\usepackage{bm}
\usepackage{epsfig} 
\usepackage{csquotes}

\usepackage[lofdepth,lotdepth]{subfig}
\begin{document}

\preprint{APS/123-QED}

\title{Machine learning for analysis of plasma driven Ion source }

\author{N. Joshi}
\email{njoshi@fias.uni-frankfurt.de}
\author{ O. Meusel}
\author{ H. Podlech}


\affiliation{%
Institute for Applied Physics, Goethe University, \\
60438 Frankfurt, Germany 
}

\date{}

\begin{abstract}

Recently, neural networks have found many applications in different fields including Genetics, Pharmacy, Astrophysics and High Energy Physics \cite{Gene_ML, Astro_ML,Baldi}.
In the field of accelerator physics it has been used for control systems \cite{Edelen}.
In this paper we present the results based on machine learning techniques motivated to predict the behaviour of ion source in terms of composition of the ion beam while using hydrogen gas to produce $H^+$ ions.
In the framework of the stellarator type Figure-8 Storage Ring (F8SR) project, a volume type ion source was designed for the low energy ion beam transport experiments.
In a first step the functioning of this ion source was studied and the results were published, but only small number of measurements were analysed as the main requirement for the on going experiment was fulfilled. 
Though at a later stage, more number of measurements were recorded with larger parameter space to investigate the properties of extracted ion beams from this source further.
With recent interests and improved techniques in the applications of  machine learning algorithms  data analysis using neural network  has been applied to study the ion beams from this ion source.

\end{abstract}

\maketitle

\section{\label{sec:level1}Introduction\protect\\
 }

Particle accelerators are hosted at many facilities ranging from research institutes to medical facilities. 
The device for ion beam production of any accelerator is always an ion source, thus it becomes clear that the performance of the ion source  would dominate the ion beam quality of the driver accelerator.
A hot filament driven gas discharge ion source is known to deliver a stable high current beam.
The properties of such an ion source are described in \cite{NJ}.
The design of the ion source is based on the models that can describe plasma properties such as density, temperature but the plasma chemistry and production mechanism of ions is not taken into account.
In case of inert gas only one type of ions are produced e.g. $He^{+}$ from Helium gas.
But in case of Hydrogen gas the description of the plasma becomes difficult  as multiple processes are involved those produce different molecular ions namely $H^{+}$, $H_2^{+}$ and $H_3^{+}$.
Thus experiments must be performed to find the optimal conditions to extract a particular ion specie.
A human operator plays an important role to tune the ion source for matching parameters.
The operator must retune the source after every machine shut down.

In order to predict the production of a particular ion species depending on the source parameters and to reduce the dependency on experience based operation a machine learning approach was investigated.
In particular, neural networks are well suited for problems that consist of large data with multi- variable parameter space.

\section{\label{sec:level1}Experimental data\protect\\
 }

As described in the paper, the composition of ion beams depends on the various source parameters like arc current, arc potential, gas pressure and magnetic field.
All these parameters determine the plasma properties such as density and temperature. 
Although the direct correlation between plasma state and operating parameters is not well established.
Only small number of measurements were analysed  in the above mentioned parameter space earlier. 
The so called $r_{occ}$ parameter defined in the paper showed that it takes high value at certain parameter set \cite{NJ} . 
In the earlier publication only a  few measurements were described. 
A full dataset was available only for $P = 1.2~ mbar$ at $V_{arc} = 80~ V$.
In subsequent experiments the parameter space was fully explored to collect a larger dataset.
Table \ref{Table1} gives the values for all the parameters used in experiments

\begin{center}
\begin{tabular}{ c c c c}
 Parameter & Range & Resolution & No. of measurements \\ 
$V_{arc}$ & $60 - 120 ~V$ & $10~ V$ &  7 \\  
$I_{arc}$ & $1- 12 ~A$ & $1~A$ &12 \\
$B~Field$ & $0 - 30 ~mT$ & $2~mT$ & 15 \\
$P$ & $1.2 -6.4 ~mbar$ & NA & 6 \\    
\end{tabular}
\label{Table1}
\end{center}

Thus in total about $7560$  measurements are now available.

\begin{figure}[hhh]
    \centering
    \includegraphics[width=8cm]{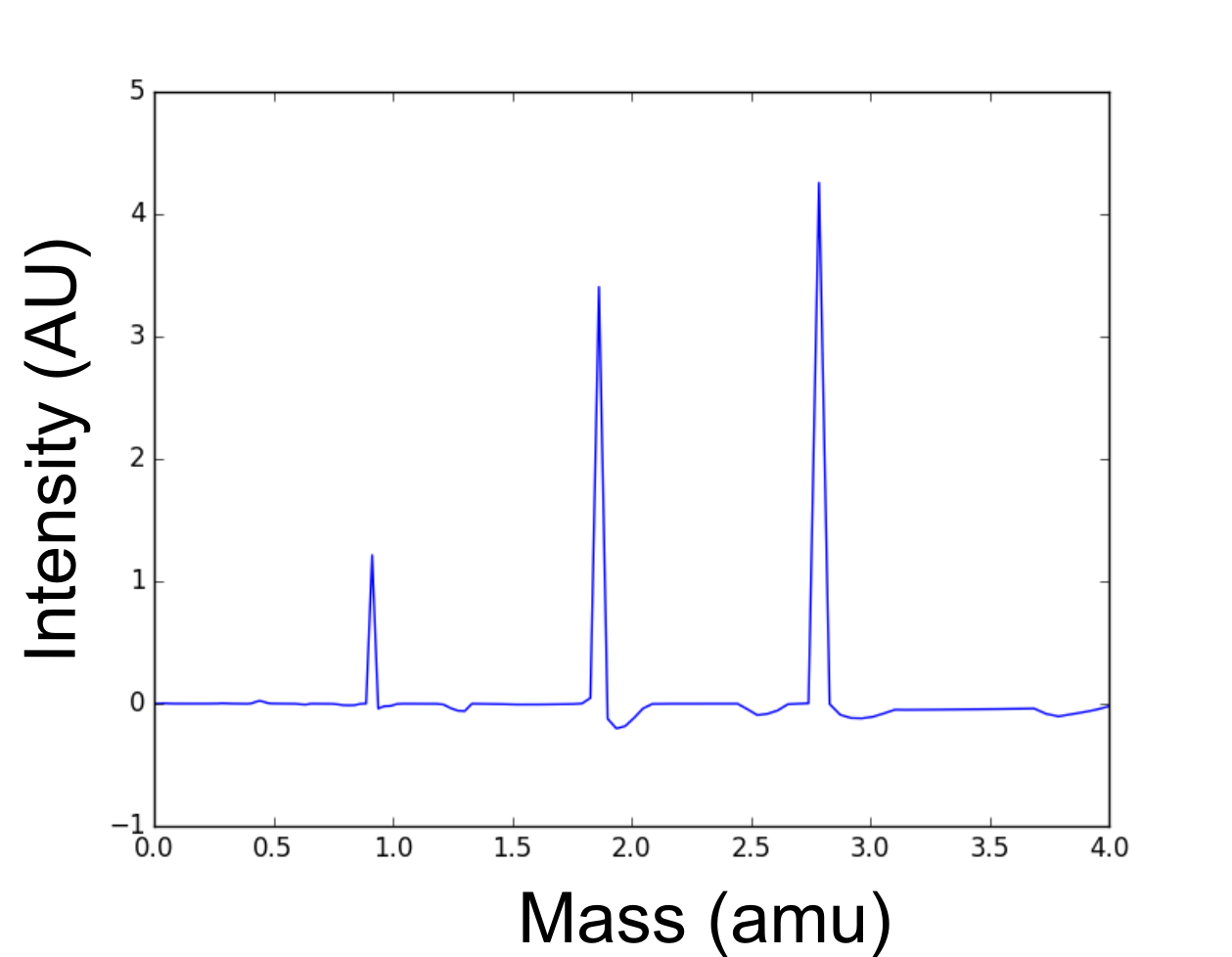}
    \caption{An example of a mass spectra showing  presence of $H^+$, $H_2^+$ and $H_3^+$ ions.
	Source parameter: $B= 4 mT$, $I_{arc}= 10 A$, $V_{arc} = 60~V$, $P = 1.2~ mbar$. Not: A slight offset on the mass axis corresponds to the energy calibration error of about 10 keV. }
    \label{spectra}
\end{figure}

A dataset was composed by extracting relative intensitites each ion type from mass spectra,  for different parameters of  $B, I_{arc},  V_{arc} , P$.
Fig. \ref{spectra} shows one such spectra showing the presence of peaks at $ m = 1,2~ and ~3$ corresponding to $H^+$, $H_2^+$ and $H_3^+$ ions respectively.
The relative composition of the beam is determined by calculating area under peak to the total area under curve.
Due to the certain physical aspect some spectra were either not clear, dominated by noise, or with multiple unexpected peaks.
Only a clear spectra was taken into consideration.
 $r_{occ}$ is then calculated which shows relative presence of particular ion specie in the beam defined in earlier publication as

\begin{equation}
r_{occ} = \frac {  \eta_{H_{}^+}}{\eta_{H_{2}^+} +\eta_{H_{3}^+} }, 
\end{equation}
 
where $\eta$ is the relative percentage at that particular set of parameters.

\begin{figure}[h]
\centering
\subfloat[Subfigure 1 list of figures text][$H^+$]{
\includegraphics[width=0.3\textwidth]{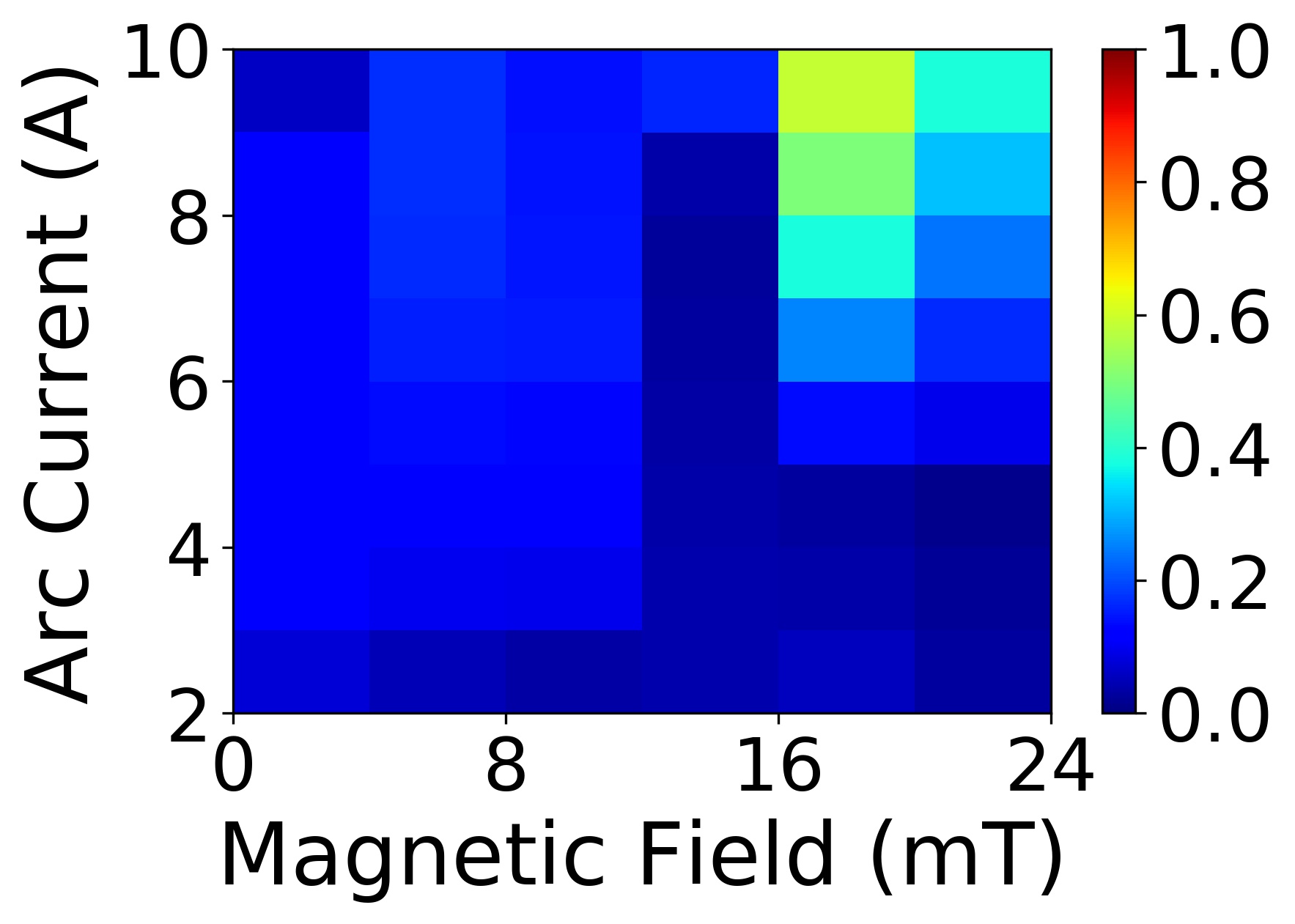}
\label{fig:subfig1}}
\qquad
\subfloat[Subfigure 2 list of figures text][$H_2^{+}$]{
\includegraphics[width=0.3\textwidth]{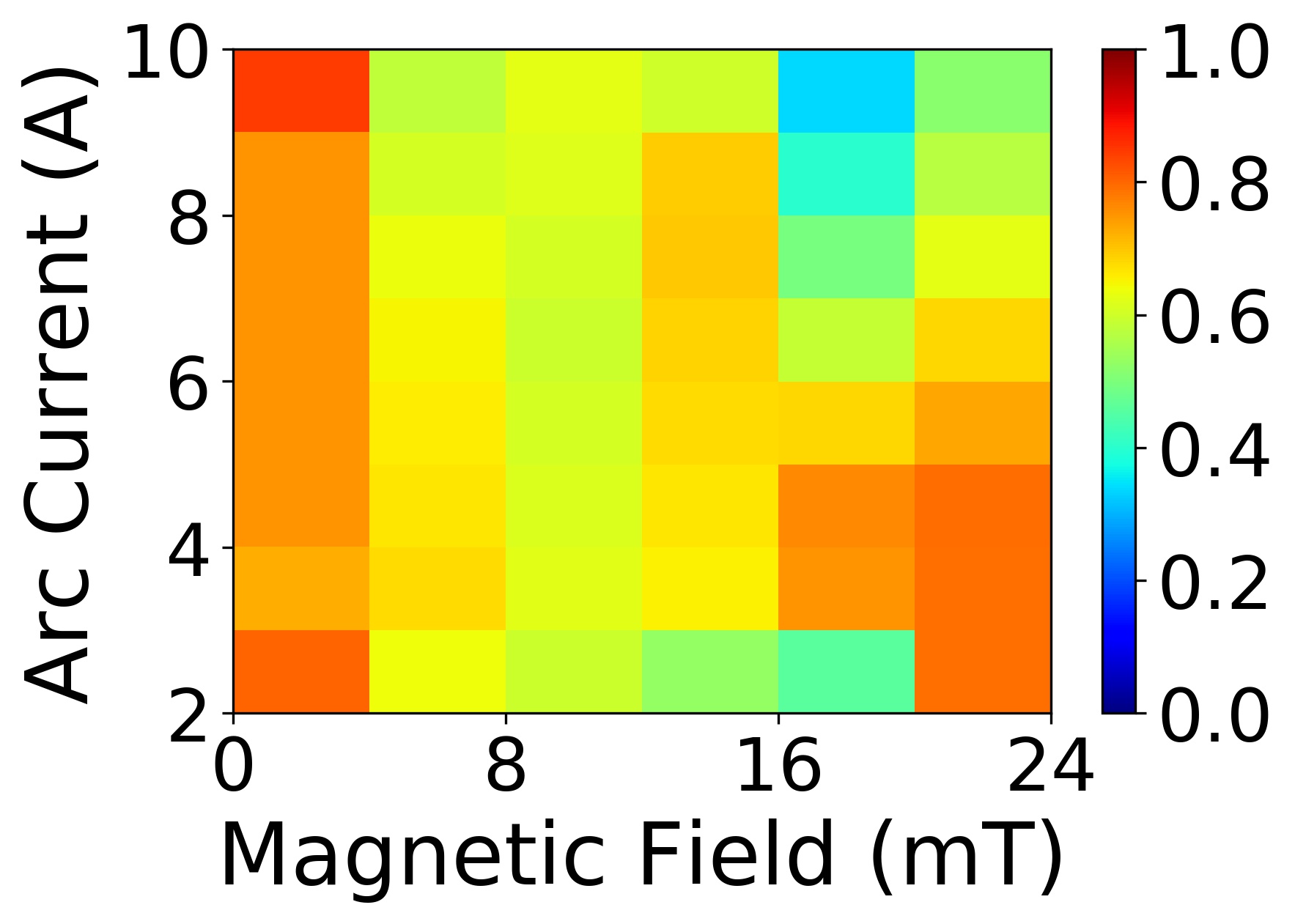}
\label{fig:subfig2}}
\subfloat[Subfigure 3 list of figures text][$H_3^{+}$]{
\includegraphics[width=0.3\textwidth]{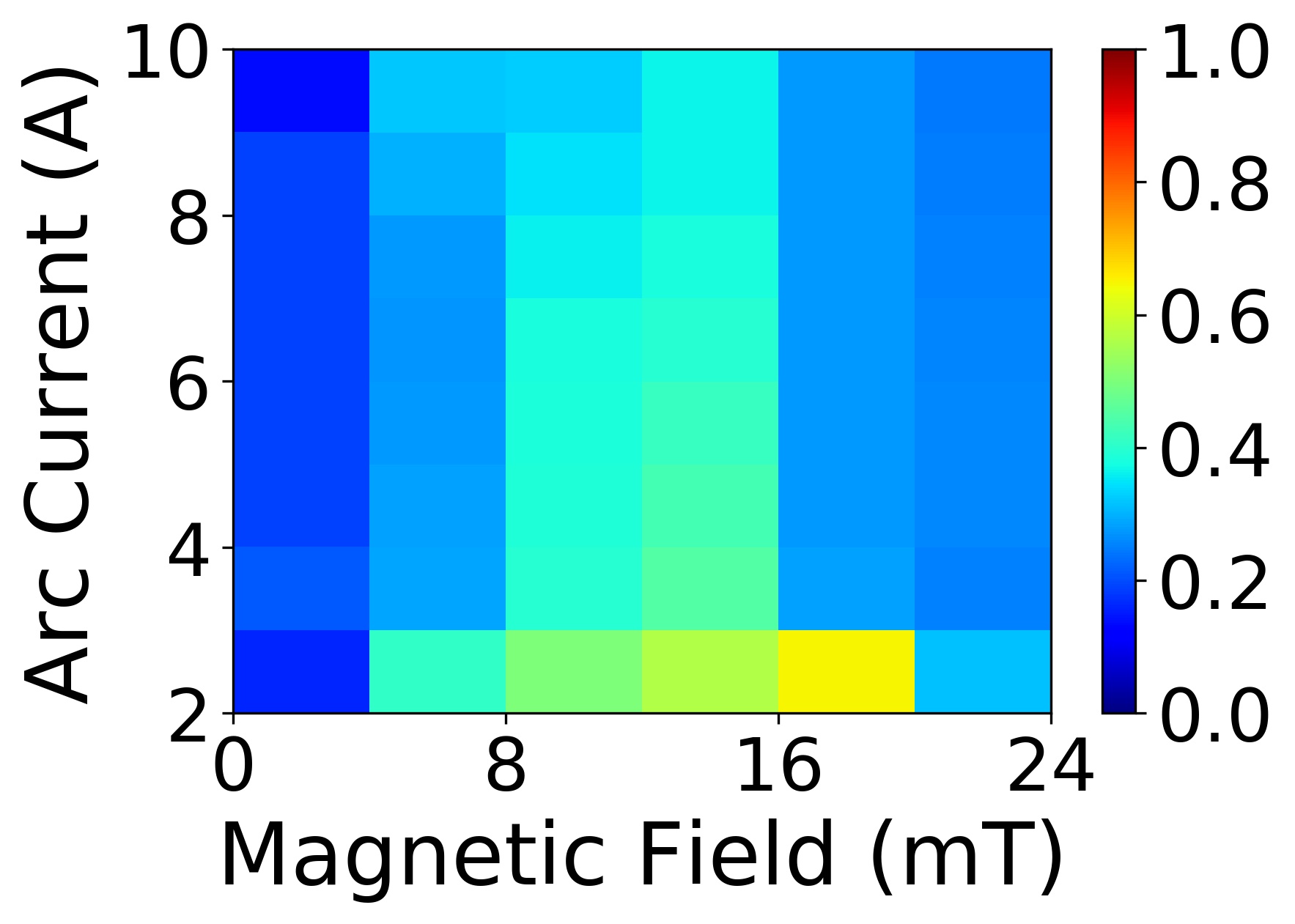}
\label{fig:subfig3}}
\qquad
\caption{$r_{occ}$ values plotted as a function of magnetic field and arc current for different ion species. $r_{occ}$ values are normalized and  colour coded for each ion species, namely (a) $H^+$, (b) $H_2^+$ and (c) $H_3^+$.
These graphs were generated by using newly accumulated measurements with higher statistics and resolution at gas pressure of $P=3.2 ~mbar$ and arc potential $V_{arc} = 100~ V$.}
\label{Fractions}
\end{figure}

Fig. \ref{Fractions} shows the  $r_{occ}$ as a function of magnetic field and arc current at $P=3.2~mbar$, $V_{arc} = 100~ V$.
A similar graph was also shown in \cite{NJ}.

\section{\label{sec:level1}Neural network based classification\protect\\
 }
 
The mechanism behind the composition of a beam in terms of operating parameters is not been studied in detail.  
The operator relies on his experience for fine tuning the ion source.
After every operation or  beam time it is expected that the operating point of the ion source moves in the parameter space.
Hence it becomes necessary to predict its behaviour from the earlier "experience".

Such type of problems are treated using neural network since quite some years but only recently used in fundamental sciences.
The learning based problems can broadly classified as \emph{ supervised} or \emph{ unsupervised}.
In this case a simple supervised classification approach was used.
We have a dataset, i.e. set of mass spectra,  containing labelled examples for training a neural network. 
A feature vector is formed from the input parameters such as arc current, voltage etc., and labels are created from composition of the ion beam.
 
It was known that the $H^+$ fraction can be maximum $55\%$, whereas $H_2^+$ and $H_3^+$ fractions can reach up to $98\%$ individually for this particular type of ion source.
The output results were classified into 4 classes according to presence of particular ions.
The class 1 corresponds to the spectra with  $H^+$ more than $30\%$. 
The class 2  corresponds to the spectra with $H_2^+$ fractions when its presence is more than $70\%$ .
The same rule applies to class 3 with $H_3^+$.
With this conditioning, the overlapping of classes is also avoided. 
The rest belongs to the class 0 or the noise. 
 
 Preparation of the data in the useful format was done using a PANDAS data frame.
 It provides data manipulation with integrated indexing; data  can be aligned and  missing data can also be handled.
 
 \begin{figure}[hhh]
    \centering
    \includegraphics[width=12cm]{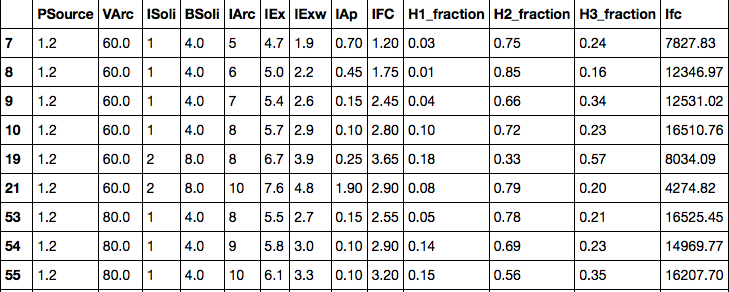}
    \caption{Screenshot of PANDAS data frame showing the organization of the dataset. 
      }
    \label{Snap_PANDAS}
\end{figure}

\begin{figure}[hhh]
    \centering
    \includegraphics[width=8cm]{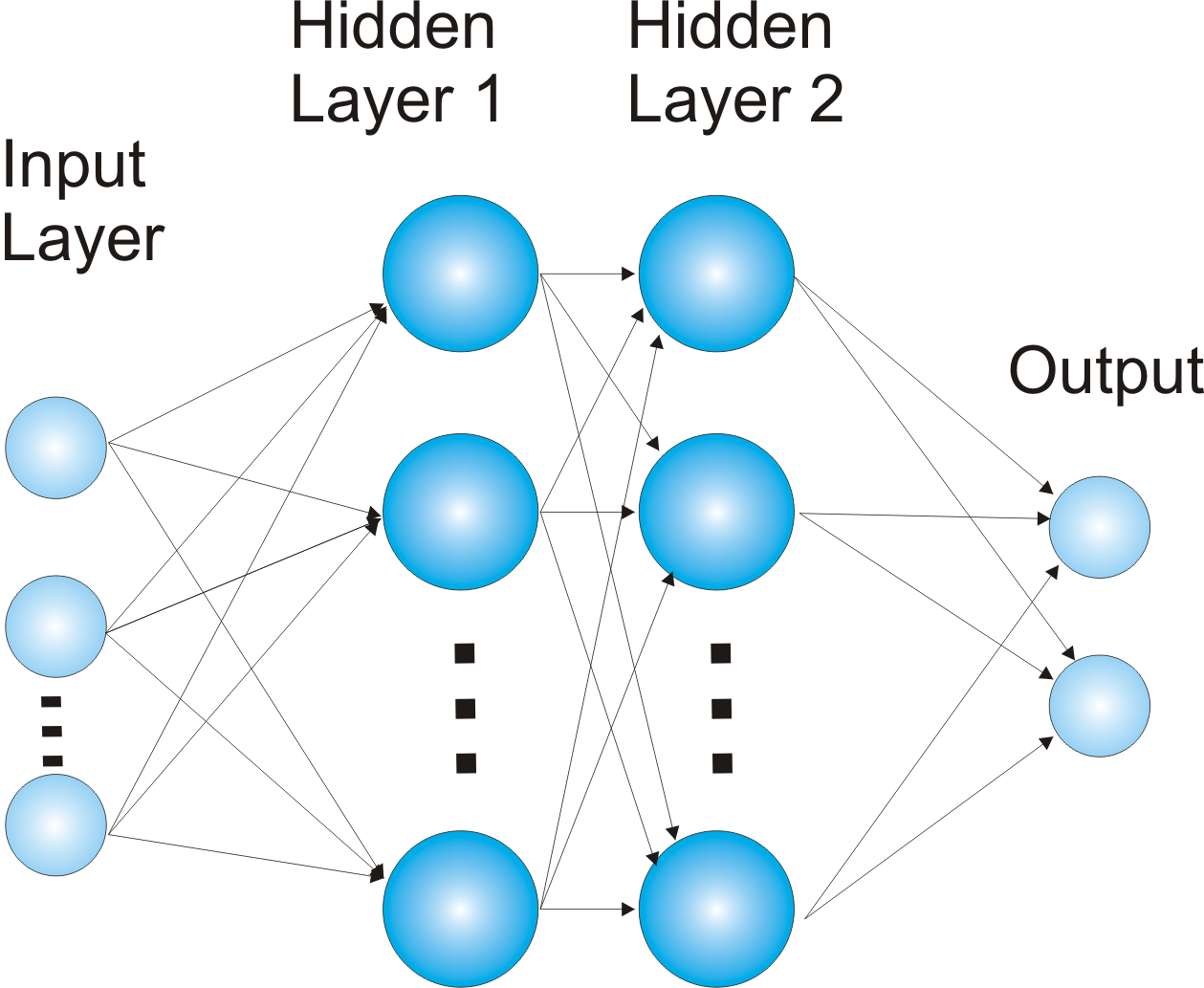}
    \caption{ Schematic representation of multi layer neural network. For this work 2 hidden layers were used with 5 units each.
      }
    \label{MLP}
\end{figure}

A multi-layered neural network was designed with 2 hidden layers, with 5 units in each layer (see Fig. \ref{MLP}).
Not much literature is available regarding the choice of the number of hidden layers and the  number of units.
In this case it was found that above mentioned numbers gives an optimal solution.

 \begin{figure}[hhh]
    \centering
    \includegraphics[width=8cm]{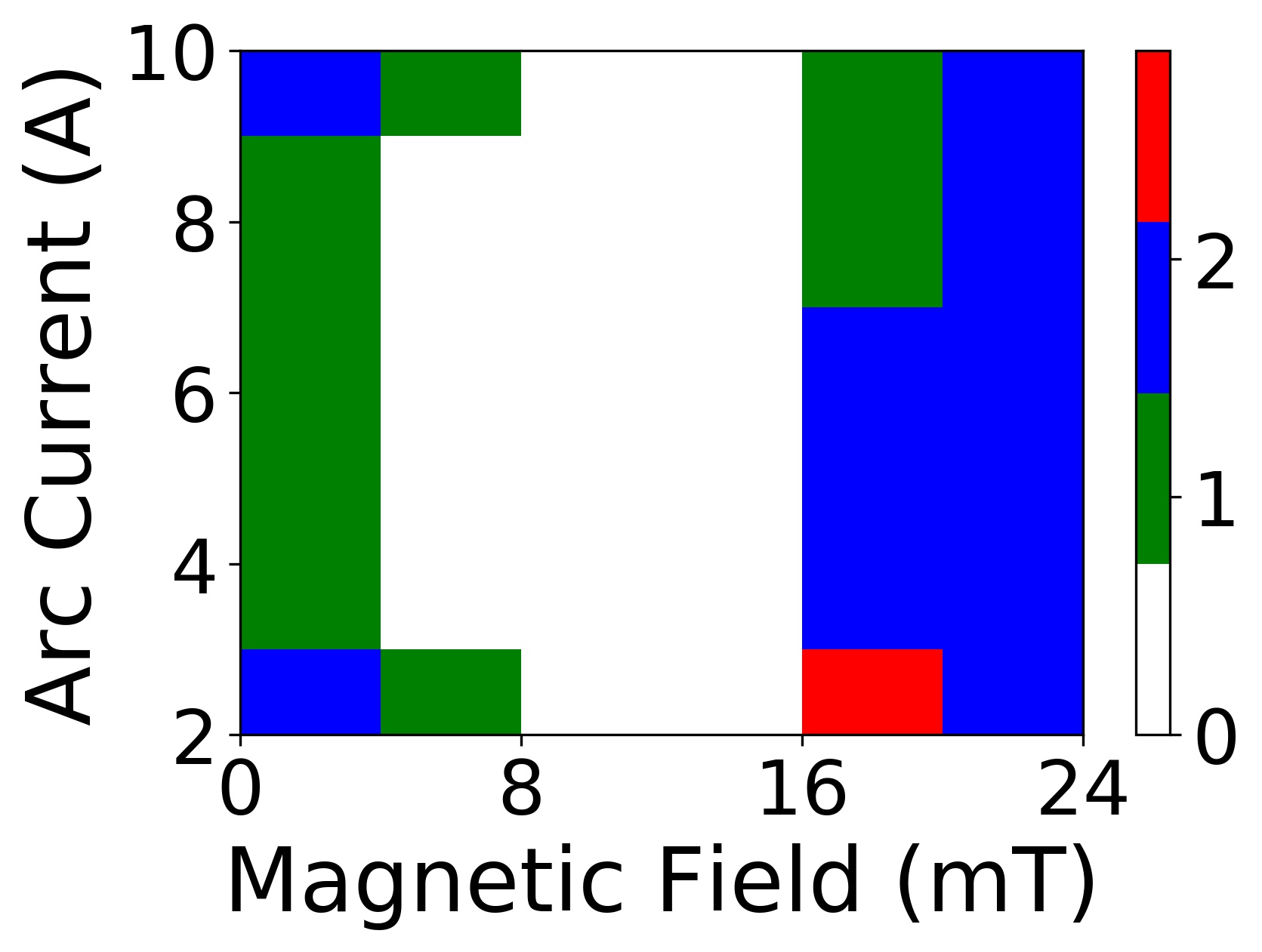}
    \caption{ Classification spitted out by neural network for a gas pressure of $P=3.2 ~mbar$,  and arc potential $V_{arc} = 100~ V$. Please compare Fig. \ref{Fractions}. Colour coding corresponds to ion beam  dominated by particular ion specie. Red: $H_3^+$, Blue:  $H_2^+$, Green $H^+$, and white space corresponds to noise. 
      }
    \label{fig:result_1}
\end{figure}

The input data was pre processed using \emph{Sklearn} package i.e. input vectors were normalised and the dataset was divided into the training and testing set randomly. 
Typically this ratio is chosen as $4:1$.
The class labels were converted into logits.
A sigmoid activation function defined as,

\begin{equation}
S(x) = \frac{1}{1 + e ^{-x}}
\end{equation}
 
was used along with the Adam optimiser for back propagation.

Accuracy was calculated separately for each ion specie by taking the ratio of correctly classified cases to total number of observations.
In case of $H_2^{+}$ and   $H_3^{+}$ approximately $86\%$ accuracy was achieved. whereas  for $H^{+}$ ions it was quite low, leading up to $56\%$.
One of the main reason was a high noise level in the data.
Fig. \ref{fig:result_1} shows predictions for ion species in the beam composition for the gas pressure value of   $P=3.2~mbar$ , and arc potential  $V_{arc} = 100 V$.
The graph depicts that  the $H_2^{+}$ and   $H_3^{+}$ dominated beams has been predicted correctly but at $I_{arc} = 4-6~A$ it has wrongly predicted $H^+$ beam.

\section {Conclusions and Outlook}
\label {Conclusions}

Although one may argue that the prediction accuracy for particular ions is not very accurate, but considering the fact that in usual cases the ion source acts like a black box, such predictions would help considerably to quickly find stable operational parameters.
The ongoing efforts involve extracting cleaner signal from back ground noise using recurrent network as autoencoders.
These algorithms are being generalized for operation and design of accelerator devices and subcomponents.
A dedicated control system is being designed base on  artificial intelligence which can perform online pattern recognition and prediction systems for autonomous operations.

\end{document}